\documentstyle{mn-1.4}

\newif\ifAMStwofonts
\AMStwofontstrue

\def\sun{\ifmmode\odot\else$\odot$\fi}
\def\H2{\hbox{H$_{2}$}}
\def\HII{\hbox{H\,{\sc ii}}}

\def\mic{$\mu$m}
\def\kms{${\rm km~s}^{-1}$}
\newcommand{\etal}{{\rm et al.}}
\newcommand{\eg}{{\it e.g.,}}
\newcommand{\ie}{{\it i.e.,}}
\newcommand{\cmt}{{$\rm cm^{-3}$}}
\newcommand{\cmtwo}{{$\rm cm^{-2}$}}
\newcommand{\sone} {{$\rm s^{-1}$}}
\newcommand{\arc}{{arcsec$^{-2}$}}

\ifoldfss
  \ifCUPmtlplainloaded \else
    \NewTextAlphabet{textbfit} {cmbxti10} {}
    \NewTextAlphabet{textbfss} {cmssbx10} {}
    \NewMathAlphabet{mathbfit} {cmbxti10} {} 
    \NewMathAlphabet{mathbfss} {cmssbx10} {} 
  \fi
  \ifAMStwofonts
    \ifCUPmtlplainloaded \else
      \NewSymbolFont{upmath} {eurm10}
      \NewSymbolFont{AMSa} {msam10}
      \NewMathSymbol{\upi}     {0}{upmath}{19}
      \NewMathSymbol{\umu}     {0}{upmath}{16}
      \NewMathSymbol{\upartial}{0}{upmath}{40}
      \NewMathSymbol{\leqslant}{3}{AMSa}{36}
      \NewMathSymbol{\geqslant}{3}{AMSa}{3E}
           \let\oldleq=\leq
           \let\oldgeq=\geq

    \fi
  \fi
\fi 

\ifnfssone
  \newmathalphabet{\mathit}
  \addtoversion{normal}{\mathit}{cmr}{m}{it}
  \addtoversion{bold}{\mathit}{cmr}{bx}{it}
  \newmathalphabet{\mathbfit} 
  \addtoversion{normal}{\mathbfit}{cmr}{bx}{it}
  \addtoversion{bold}{\mathbfit}{cmr}{bx}{it}
  \newmathalphabet{\mathbfss} 
  \addtoversion{normal}{\mathbfss}{cmss}{bx}{n}
  \addtoversion{bold}{\mathbfss}{cmss}{bx}{n}
  \ifAMStwofonts
    \ifCUPmtlplainloaded \else
      \UseAMStwoboldmath
      \makeatletter
      \new@mathgroup\upmath@group
      \define@mathgroup\mv@normal\upmath@group{eur}{m}{n}
      \define@mathgroup\mv@bold\upmath@group{eur}{b}{n}
      \edef\UPM{\hexnumber\upmath@group}
      \new@mathgroup\amsa@group
      \define@mathgroup\mv@normal\amsa@group{msa}{m}{n}
      \define@mathgroup\mv@bold\amsa@group{msa}{m}{n}
      \edef\AMSa{\hexnumber\amsa@group}
      \makeatother
      \mathchardef\upi="0\UPM19
      \mathchardef\umu="0\UPM16
      \mathchardef\upartial="0\UPM40
      \mathchardef\leqslant="3\AMSa36
      \mathchardef\geqslant="3\AMSa3E
           \let\oldleq=\leq
           \let\oldgeq=\geq

    \fi
  \fi
\fi 

\ifnfsstwo
  \DeclareMathAlphabet{\mathbfit}{OT1}{cmr}{bx}{it}
  \SetMathAlphabet\mathbfit{bold}{OT1}{cmr}{bx}{it}
  \DeclareMathAlphabet{\mathbfss}{OT1}{cmss}{bx}{n}
  \SetMathAlphabet\mathbfss{bold}{OT1}{cmss}{bx}{n}
  \ifAMStwofonts
    \ifCUPmtlplainloaded \else
      \DeclareSymbolFont{UPM}{U}{eur}{m}{n}
      \SetSymbolFont{UPM}{bold}{U}{eur}{b}{n}
      \DeclareSymbolFont{AMSa}{U}{msa}{m}{n}
      \DeclareMathSymbol{\upi}{0}{UPM}{"19}
      \DeclareMathSymbol{\umu}{0}{UPM}{"16}
      \DeclareMathSymbol{\upartial}{0}{UPM}{"40}
      \DeclareMathSymbol{\leqslant}{3}{AMSa}{"36}
      \DeclareMathSymbol{\geqslant}{3}{AMSa}{"3E}
           \let\oldleq=\leq
           \let\oldgeq=\geq

    \fi
  \fi
\fi 

\ifCUPmtlplainloaded \else
  \ifAMStwofonts \else 
    \def\upi{\pi}
    \def\umu{\mu}
    \def\upartial{\partial}
  \fi
\fi

\title[Molecular Hydrogen in Parsamyan~18]
       {Molecular Hydrogen Line Emission from the Reflection Nebula
       Parsamyan~18}
\author[S. Ryder et~al.]
       {S. D. Ryder\thanks{Present address: Joint Astronomy
        Centre, 660 N. A'Ohoku Place, Hilo, HI 96720, U.S.A. E-mail:
        sryder@jach.hawaii.edu.},
        L. E.~Allen, M. G. Burton, M. C. B. Ashley and J. W. V. Storey \\
        School of Physics, University of New South Wales, Sydney 2052,
        Australia}
        \date{Accepted 1997 September 15.
      Received August 7;
      in original form 1997 June 5}

\pagerange{\pageref{firstpage}--\pageref{lastpage}}
\pubyear{1997}

\begin{document}

\maketitle

\label{firstpage}

\begin{abstract}

The newly-commissioned University of New South Wales Infrared
Fabry-Perot (UNSWIRF) has been used to image molecular hydrogen
emission at 2.12 and 2.25~\mic\ in the reflection nebula Parsamyan~18.
P~18 is known to exhibit low values of the $(1-0)/(2-1)~S(1)$ ratio
suggestive of UV-pumped fluorescence rather than thermal excitation by
shocks. Our line ratio mapping reveals the full extent of this
fluorescent emission from extended arc-like features, as well as a
more concentrated thermal component in regions closer to the central
exciting star. We show that the emission morphology, line fluxes, and
gas density are consistent with the predictions of photodissociation
region (PDR) theory. Those regions with the highest intrinsic
$1-0~S(1)$ intensities also tend to show the highest
$(1-0)/(2-1)~S(1)$ line ratios. Furthermore, variations in the line
ratio can be attributed to intrinsic fluctuations in the incident
radiation field and/or the gas density, through the self-shielding
action of \H2. An isolated knot of emission discovered just outside
P~18, and having both an unusually high $(1-0)/(2-1)~S(1)$ ratio and
relative velocity provides additional evidence for an outflow source
associated with P~18.
\end{abstract}


\section[]{Introduction}

The `cometary nebula' Parsamyan 18 (P~18 = NGC~2316 = L~1654) has
properties characteristic of both reflection and emission
nebulae. Near-infrared emission lines of molecular hydrogen have been
observed, as well as emission features at 3.3, 6.2, 7.7, 8.6, and
11.3~\mic\ (Sellgren 1986; Burton et~al. 1990; Sellgren, Werner, \&
Allamandola 1996) attributed to aromatic hydrocarbons. Similar
emission features and near-infrared colours were found in over half of
the visual reflection nebulae surveyed by Sellgren
et~al. (1996). L\'{o}pez et~al. (1988) used a combination of optical,
infrared, and radio continuum observations to argue that the two
brightest regions of P~18 are not in fact stars, but actually
represent light from a heavily-obscured \HII~region breaking through a
dust shell surrounding the exciting early B-type star.
On the basis of (unpublished) optical spectra however, Sellgren et~al. (1996)
list both components as being stellar, with the northeastern
star (Star `A', a B2-3e star with $V=13.21$) illuminating the dust as
well as exciting the atomic and molecular hydrogen.

P~18 was only the second object (after NGC~2023; Gatley \etal\ 1987)
found to have infrared \H2~emission line ratios similar to the
predictions of UV-pumped fluorescence, rather than thermal excitation
by shocks (Sellgren 1986).
In nearly all such
studies to date, line ratios and inferred excitation mechanisms are
based on large-aperture spectra, or from only a single pointing. With the
advent of imaging infrared Fabry-Perot systems (with resolving powers
$R\ga1000$ that provide the necessary contrast between line and
continuum), it has now become possible to carry out line ratio {\em
mapping\/}. This could be a useful aid in discriminating between the
sources of excitation, on the basis of line ratio trends and
morphology. In this paper, we report the results of our \H2~line
imaging of P~18 using the University of New South Wales Infrared
Fabry-Perot (UNSWIRF).


Our results show that `collisional fluorescence' is occurring in
P~18. That is, pure fluorescent line ratios are thermally modified
by the effects of collisions.  This occurs in dense molecular clouds
where H$_{2}$ self-shielding maintains a significant column density of
the molecule close to the cloud surface (\ie\ A$_{V} < 1$), where the gas
temperature can reach a few thousand degrees (\eg\ Sternberg \&
Dalgarno 1989; Burton, Hollenbach, \& Tielens 1990 [hereafter BHT]).
Emission line ratios are then sensitive to the density and
radiation field, and so observations with sufficiently high spatial
resolution can be used to trace the variation of these parameters
within a source.  Large aperture measurements of both H$_{2}$ and
high--$J$ CO lines (\eg\ Hayashi \etal\ 1985; Stutzki \etal\ 1990) of
many such `photodissociation regions' or PDRs (Tielens \& Hollenbach
1985; BHT) have indicated they are clumpy, containing regions where
the density must be ten to a hundred times greater than the
surrounding cloud.  High spatial resolution H$_{2}$ line ratio
measurements may now be used to probe how this high density gas is
distributed.

\section[]{Observations}\label{s:obs}

UNSWIRF is a 70~mm diameter Queensgate (UK) Ltd. etalon with
$R\sim4000$, tunable over both the $H$ and $K$ windows (Ryder et~al. 1997).
When used in conjunction with the wide-field mode of
IRIS\footnote{The Infrared Imager and Spectrometer (Allen et~al. 1993)
uses a $128\times128$ HgCdTe array manufactured by Rockwell
International Science Centre, CA.} at the $f/36$ focus of the
Anglo-Australian Telescope, a (roughly circular) field of 1.7~arcmin
diameter at 0.77~arcsec~pixel$^{-1}$ is produced. Appropriate
1~per~cent bandpass filters are used to ensure only a single order is
passed by the etalon.

Observations of P~18 in the \H2~$1-0~S(1)$ (2.122~\mic) and
\H2~$2-1~S(1)$ (2.248~\mic) lines were obtained on 1996 April~4 UT.
A sequence of four images spanning the peak of the 2.122~\mic\ line
(as determined from observations of OMC-1) and separated by $\sim2/3$
of the instrumental profile width were obtained with UNSWIRF, while
for the 2.248~\mic\ line, just three images spaced $\sim1/2$ of the
profile width apart were taken. In each case, another image at
3--4~profile widths from the line centre was obtained in order to
sample the continuum. An integration time of 120~s per etalon setting
was used, with non-destructive readouts every 5~s. The same exposure
sequence was carried out on the sky 5~arcmin east of P~18. The object
and sky exposures were repeated (with slight offsets from the previous
positions), leading to total on-source integration times of 1920~s and
2880~s in the 2.122 and 2.248~\mic\ lines, respectively.

Dark current subtraction and linearisation are performed during
readout. The object frames were sky-subtracted, and then flatfielded
using normalised dome flatfields, using in each case images at the
matching etalon settings. All frames were first registered using field
stars, and all frames at a given etalon setting were averaged
together. The continuum frames, appropriately scaled, were then
subtracted from all other images to leave just pure line emission in
each.

At this stage, the emission line images were `stacked' into cubes of
increasing etalon spacing / wavelength, so that a Lorentzian profile
could be fitted to the (3 or 4) point spectrum at each pixel.  Having
already fixed the continuum level to be zero, the number of free
parameters in the fitting was further reduced by constraining the
width of the profile to be the same as that found by fits to
high-resolution scans of Krypton arc lines (2.1165~\mic\ and
2.2485~\mic\ for the 2.12~\mic\ and 2.25~\mic\ cubes, respectively). We
justify this on the basis that the narrow \H2\ line widths
(16--20~\kms) observed in P~18 by Burton et~al. (1990) would not be
resolvable by UNSWIRF (${\rm FWHM}\sim65$~\kms\ at 2.12~\mic). The
output of the line fitting includes maps of the peak intensity, as
well as the etalon setting corresponding to that peak. Blanking masks
are constructed by requiring
that the fitted peak position must lie within the range covered by the etalon
sequence. Flux calibration, and continuum scaling factors, are
provided by photometry of $10\times1$~s observations of the
spectroscopic standard BS~2882 ($K=5.18$), also at the same etalon
settings used for P~18.

\section[]{Results}\label{s:res}

\begin{figure}
\vspace{7cm}
\includegraphics{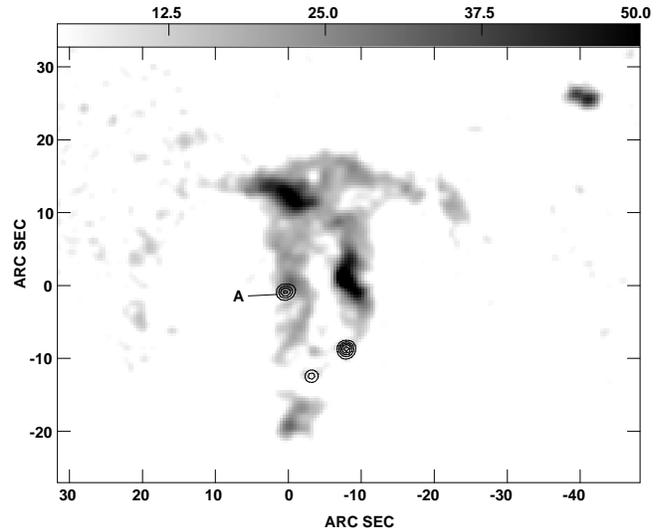}
\caption{Grey-scale image of the \H2~$1-0~S(1)$ (2.122~\mic) emission
line intensity in P~18, with contours of the 2.12~\mic\ continuum
sources overlaid. The continuum source identified previously as star
`A', and assumed to be the source of excitation of the \H2, is marked.
The coordinate system shown is in arcseconds north and east of star
`A', which has coordinates of $06^{\rm h} 59^{\rm m} 41\fs6$,
$-07\degr 46\arcmin 29\arcsec$ (J2000). The grey scale-wedge indicates
flux densities in units of $10^{-16}$~ergs~cm$^{-2}$~s$^{-1}$~arcsec$^{-2}$.
The 0.77~arcsec pixels have been re-sampled onto a finer grid, then smoothed
with a 1~arcsec Gaussian to accentuate the overlapping arc features mentioned
in the text. The increased noise in the northeast quadrant is due to a
defect in one of the etalon anti-reflection coatings.}
\label{f:212}
\end{figure}

\begin{figure}
\vspace{7cm}
\includegraphics{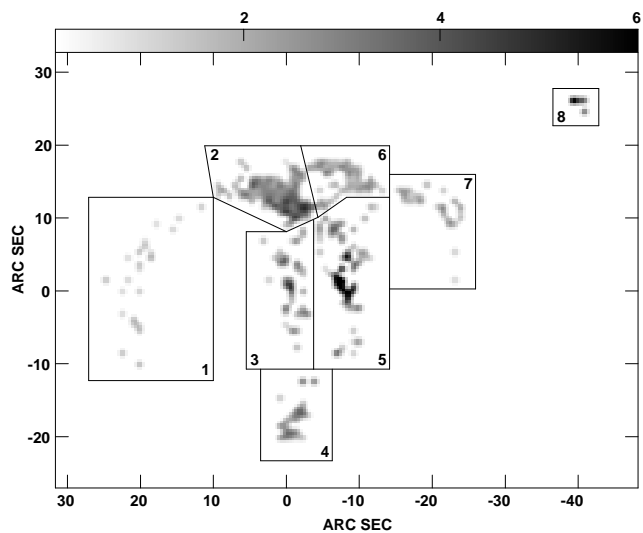}
\caption{Grey-scale image of the \H2~$1-0~S(1)/2-1~S(1)$ ratio in
P~18, for all points in which a reliable detection $({\rm S/N} > 3)$
at 2.25~\mic\ was achieved. The numbered polygons outline the eight
separate emission regions in and around P~18, as given in
Table~{\protect{\ref{t:ratio}}}.  As in Figure~\protect{\ref{f:212}},
the 0.77~arcsec pixels have been re-sampled onto a finer grid, but no
smoothing has been applied.}
\label{f:h2r}
\end{figure}

Figure~\ref{f:212} shows a grey-scale map of the 2.12~\mic\ line
intensity, with contours of the three brightest 2.12~\mic\ continuum
sources superimposed (including star `A') to aid in comparing our new
data to that elsewhere in the literature (e.g., L\'{o}pez et~al. 1988).
The H$_{2}$ emission
structure can be conveniently divided into a number of distinct
regions, as outlined in Figure~\ref{f:h2r}. The emission line
fluxes of each region are tabulated in Table~\ref{t:ratio}.
The integrated emission line flux of P~18 measured from the 2.12~\mic\
image is $(13.1\pm0.6)\times10^{-13}$~ergs~cm$^{-2}$~s$^{-1}$,
compared with the $(4.5\pm0.3)\times10^{-13}$~ergs~cm$^{-2}$~s$^{-1}$
found in the matching 2.25~\mic\ image.
Uncertainties in the absolute flux calibration and line-fitting procedures
are roughly comparable in magnitude with the noise contribution
($\sim1.8\times10^{-16}$~ergs~cm$^{-2}$~s$^{-1}$) per pixel over
which the emission is summed.

\begin{table*}
 \caption{H$_{2}$ Line Intensities, Ratios and Velocities in P~18}
 \label{t:ratio}
 \begin{tabular}{lrrccrcccr}
\hline
Region$^a$ & d$^b$ & FUV & $1-0~S(1)$ & $2-1~S(1)$ & N$^d$ &
Ratio & Peak $(1-0)~S(1)$  & Ratio & Relative \\
       &              & Field$^c$   & Flux       & Flux       &    &
      & Flux in region & at Peak   & Velocity$^e$     \\
       & (arcsec)     & (G$_0$) & (ergs \cmtwo \sone) & (ergs \cmtwo \sone) & &
      & (ergs \cmtwo \sone arcsec$^{-2}$) & & (\kms) \\
\hline
Whole Source &\ldots& \ldots & $13.1\pm0.6 (-13)$  & $4.5\pm0.3 (-13)$ & 2147 &
 $2.9\pm0.4$ & \ldots & \ldots & \ldots \\
1 (E Arc)      & 22   & 2400   & $2.7\pm0.1 (-13)$ & $1.1\pm0.1 (-13)$ &  766 &
 $2.4\pm0.3$ & $1.7 (-15)$ & 1.6  & $+8$ \\
2 (N Peak)     & 15   & 5060   & $2.2\pm0.1 (-13)$ & $8.1\pm0.5 (-14)$ &  248 &
 $2.7\pm0.3$ & $5.5 (-15)$ & 2.9  & $+2$ \\
3 (Central Arc)&\ldots&\ldots  & $1.9\pm0.1 (-13)$ & $4.9\pm0.4 (-14)$ &  260 &
 $3.9\pm0.5$ & $3.1 (-15)$ & 4.8  & 0 \\
4 (S Peak)     & 18   & 3500   & $8.5\pm0.5 (-14)$ & $3.3\pm0.3 (-14)$ &  147 &
 $2.6\pm0.4$ & $3.1 (-15)$ & 3.3  & $-2$ \\
5 (W Peak)     &  8  & 17800   & $3.0\pm0.1 (-13)$ & $8.3\pm0.6 (-14)$ &  373 &
 $3.6\pm0.4$ & $7.9 (-15)$ & 7.0  & $-1$ \\
6 (NW Arc)     & 19  &   3200  & $1.1\pm0.1 (-13)$ & $4.6\pm0.3 (-14)$ &  141 &
 $2.5\pm0.3$ & $2.5 (-15)$ & 2.9  & $+4$ \\
7 (W Arc)      & 25  &   1800  & $9.7\pm0.5 (-14)$ & $4.0\pm0.3 (-14)$ &  172 &
 $2.4\pm0.3$ & $2.0 (-15)$ & 2.4  & $-3$ \\ 
8 (P~18-NW)$^f$ & 48 &     500 & $3.8\pm0.2 (-14)$ & $4.4\pm1.2 (-15)$ &   40 &
 $8.6\pm2.9$ & $4.9 (-15)$ & 13.7 & $+14$ \\
\hline
\end{tabular}
\medskip
~~\\
\begin{flushleft}
$^a$Numbering as in Figure~\protect{\ref{f:h2r}}.\\
$^b$Projected distance from Star A.\\
$^c$In units of the average interstellar field, for the projected distance
from Star A. \\
$^d$Number of pixels, of size $0.77'' \times 0.77''$, included in region. \\
$^e$Relative velocity of H$_{2}$ emission from region. \\
$^f$This region is not linked with the others and is probably shock-excited,
rather than irradiated by Star A (Section~\protect{\ref{s:bob}}).\\
\end{flushleft}
\end{table*}

The morphology of the infrared molecular hydrogen emission differs
somewhat from the `cometary' nature displayed in the optical (Cohen
1974), and from the broad-band infrared and radio continuum images of
L\'{o}pez et~al. (1988). Instead of a cone, the appearance is one of
two incomplete, but overlapping arcs. With reference to
Figure~\ref{f:h2r}, one arc is outlined by the contiguous regions (1,
2, 5, 4), while the other is defined by the regions (4, 3, 6, 7). Each
arc, if it were part of a complete ring, would have a diameter of
$\sim30$~arcsec, or 0.16~pc, assuming P~18 to be 1.1~kpc away (Hilton
\& Lahulla 1995).  Star `A', assumed to be the exciting source, sits
on the periphery of the arc that extends to the west, and halfway
between the centre and the edge of the easterly arc. The emission
peaks (in both the 2.12~\mic\ and the 2.25~\mic\ images) in the areas
due north and south of star `A', where these two arcs overlap. In
addition, more prominent emission at 2.12~\mic\ is found on the
western edge of the eastern arc, $\sim8$~arcsec due west of star `A',
as well as in an isolated knot of emission some 50~arcsec to the
northwest of star `A'. This isolated patch (hereafter designated
P~18-NW) has no continuum counterpart or obvious exciting source.
Although none of this emission-line structure is apparent in the
$K$-band image of L\'{o}pez et~al. (1988), it is faintly visible in
the FITS version of the $K'$ imaging survey of CO molecular outflow
sources carried out by Hodapp (1994).

Figure~\ref{f:h2r} is a grey-scale image of the $(1-0)/(2-1)$ ratio in P~18
{\em for all points in which the $2-1$ intensity could be determined}. The
values found range from $1-2$ in the eastern arc, to as much as 7 in
the nebular arc just west of Star `A', and even higher in P~18-NW
(though the $2-1$ detection here is barely at the 1~$\sigma$
level). Figure~\ref{f:h2r} also demarcates the eight distinct regions
whose fluxes, flux ratios, and relative velocities are tabulated in
Table~\ref{t:ratio}.

%
%





\section[]{Discussion}\label{s:disc}

\subsection[]{Fluorescent Molecular Hydrogen}\label{s:fluor}

The average ratio of the $1-0$ and $2-1~S(1)$ lines over the whole
of P~18 is $2.9\pm0.4$ (Table~\ref{t:ratio}), as
expected for fluorescently excited gas slightly modified by a thermal
contribution to the ${\rm v}=1-0~S(1)$ line. Pure fluorescent emission
produces a line ratio of $\sim 1.7$ (\eg\ Black \& Dalgarno 1976;
Black \& van Dishoeck 1987), but this ratio can be exceeded when some
of the gas is hot enough for the lower levels to be thermalised. We
thus conclude, as did Sellgren (1986), that molecular gas in P~18 is
irradiated by far--UV photons. It is situated in two arcs around star
`A', where some of the gas is hot enough for the ${\rm v}=1$ level to be
collisionally populated.  This is characteristic of a dense
photodissociation region (Sternberg \& Dalgarno 1989; BHT). In
these PDRs the gas is sufficiently dense that self-shielding of H$_{2}$
occurs for optical depths $ < 1$ from the cloud surface, so that
molecules can exist in this warm, primarily atomic region.
The typical $1-0~S(1)$ line flux density in P~18 is $\sim 4 \times
10^{-15}$~ergs~\cmtwo~\sone~\arc, and peaks at 8 $\times
10^{-15}$~ergs~\cmtwo~\sone~\arc\ just west of star `A'.
These fluxes are consistent with the
predictions for PDR models for gas of average density, $n \sim
10^4$~\cmt\ irradiated by far--UV photons from star `A', as we show
below.

We estimate the far--UV flux from star `A', a B2--3e star at 1.1~kpc
(Sellgren \etal\ 1996) to be 5060~G$_{0}$ in a shell $15''$ away. Here
G$_{0} = 1$ represents the average interstellar radiation field ($=1.6
\times 10^{-3}$~ergs~\cmtwo~\sone; Habing 1968). For gas of density
$\sim 10^4$~\cmt\ exposed to far--UV fields G$_{0} \sim 10^{3-5}$, BHT
predict $1-0~S(1)$ line fluxes of $\sim 3 \times
10^{-15}$~ergs~\cmtwo~\sone~\arc\ for a face-on PDR, consistent with
the data. Note that our discovery of these arcs, and our
interpretation of them as PDRs irradiated by star `A', is at odds with
the conclusion of L\'{o}pez et~al. (1988) that the continuum sources
are not stellar, and merely unobscured sight-lines towards a compact
\HII~region. While P~18 shares some of the properties of an
\HII~region, we feel it would be too much of a coincidence if one of
these spots also had a location and spectrum consistent with being the
excitation source of the PDR.

For such conditions the models predict a purely fluorescent
$(1-0)/(2-1)~S(1)$ line ratio of $\sim 1.7$, whereas the average ratio for
the source is 2.9.  This suggests that about half the $1-0~S(1)$
emission is thermally contributed, and we discuss the implications of
this further in \S~\ref{s:ratios}.

\subsection[]{Geometry and Kinematics}\label{s:geom}

The morphology suggests that star `A' is surrounded by an (incomplete)
shell seen in projection, and in this section we investigate whether
the distribution of the H$_{2}$ line emission is consistent with such
a geometry.

\begin{figure}
\vskip 15cm
\includegraphics{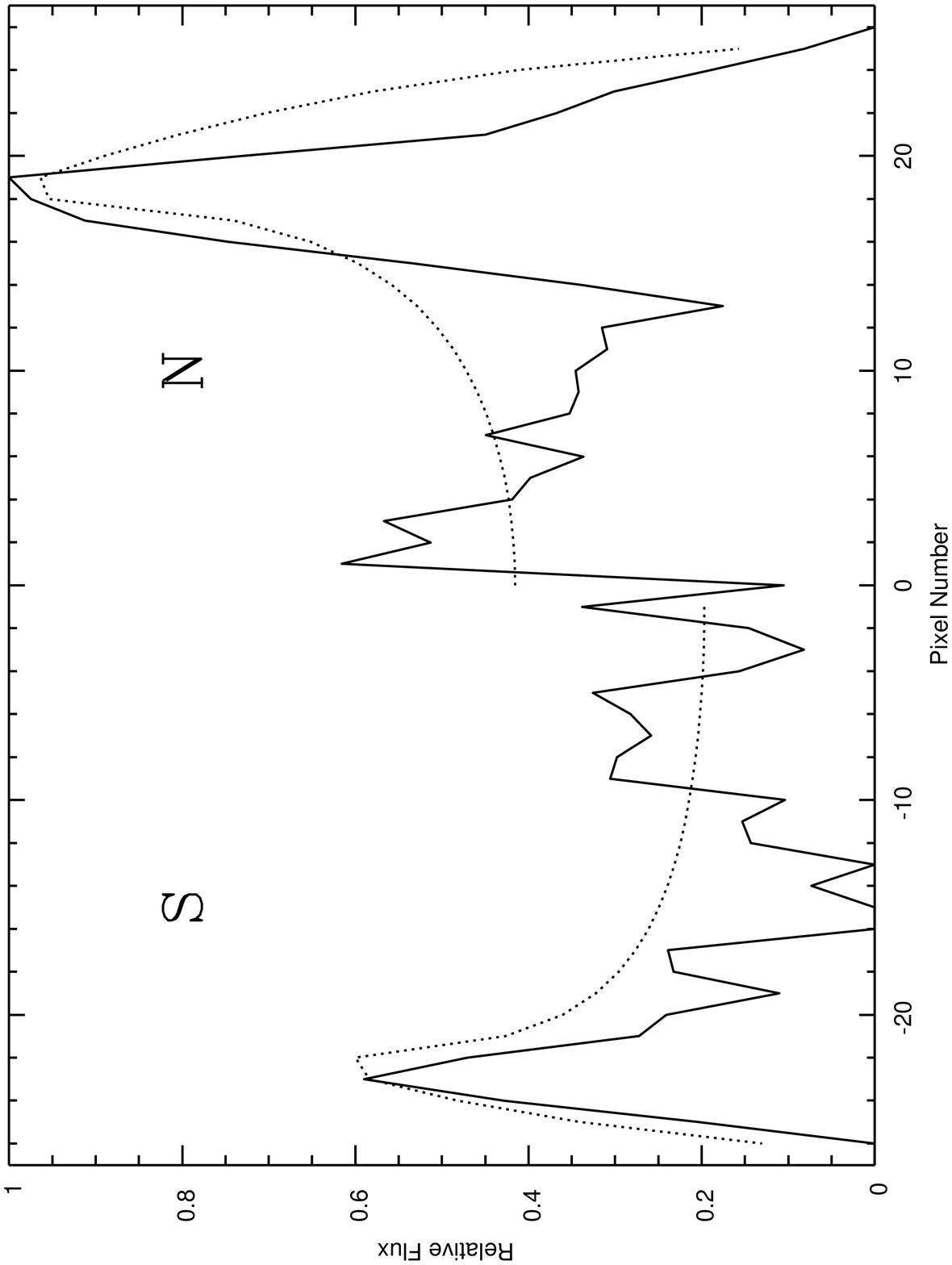}
\includegraphics{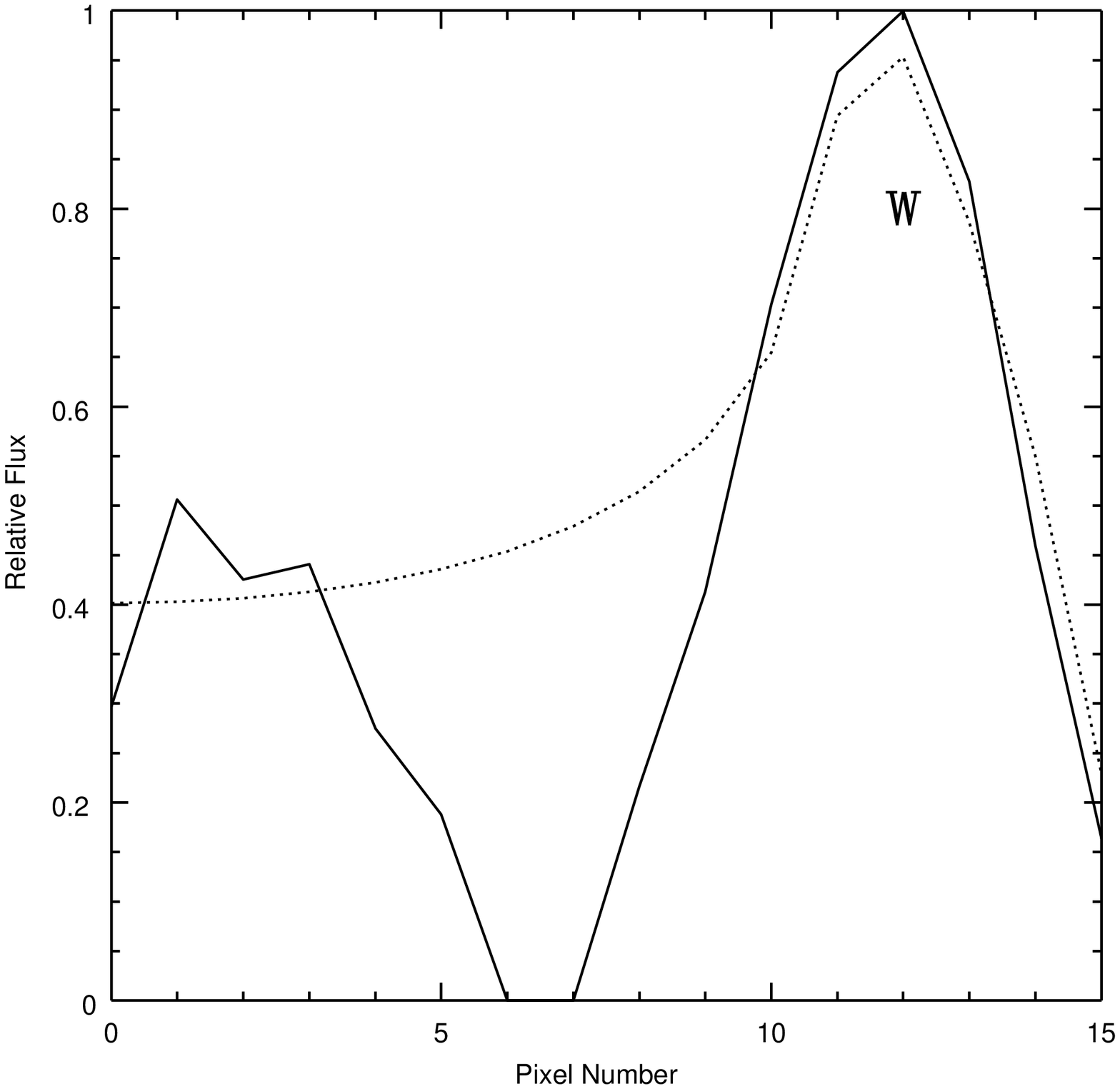}
\caption{Cuts across the 2.12~\mic\ emission in a line ({\em top}) N-S
through star `A', and ({\em bottom}) E-W through star `A' (solid lines),
together with the best fitting model results (dashed lines)
for a spherical molecular shell irradiated by star `A' with
thickness 25\% of the shell radius, where dust extinction limits
the region of \H2~excitation to A$_{V}=1$. Star `A' is at pixel (0,0)
in these plots, with 1~pixel$=1.2 \times 10^{16}$~cm for a distance
to the source of 1.1~kpc.}
\label{f:model}
\end{figure}

Consider a spherical molecular shell surrounding an illuminating star
as representing the arc through the N, W, and S peaks.  Assuming that
the average density in the cloud is low enough so that the emission is
mainly fluorescent (see \S~\ref{s:ratios}), then the H$_{2}$ emits
primarily from a sheath whose thickness is A$_{V} \sim 1$, with dust
extinction determining the depth to which the far--UV photons
penetrate.  Viewed from afar the relative intensity along a line of
sight is then proportional to the length of the sight line through the
shell.  We examined models where the thickness of the shell is varied
as a proportion of its radius (see Figure~\ref{f:model}), and found
that when this is $\sim 25\%$ a reasonable fit to the data is
achieved.  Given the source distance of 1.1~kpc this provides a
physical size to the shell, and assuming a standard extinction curve
(\ie\ $\rm A_{V} = 1 \equiv N({\rm H}_{2}) = 2 \times
10^{21}$~\cmtwo), yields an average density; specifically, for the N,
S and W peaks respectively, we obtain thicknesses of $\sim 9$, 5, and
$5\times 10^{16}$~cm for the shell and average densities of $\sim2$,
4, and $4\times 10^4$~\cmt.  The maximum filling factor in this model
(ratio of the length of our sight line through the edge of the shell
to that if seen face-on) is $\sim 3$ for a shell thickness $\sim 25\%$
of the radius. These densities are consistent with those derived by
applying PDR models to the observed line intensities.

Although the ability of UNSWIRF to resolve lines is limited by its
instrumental profile width ($\sim60 - 70$~\kms, depending on
parallelism), good signal-to-noise data such as the 2.12~\mic\
observations presented here allow us to measure changes in velocity
over the image of only a few \kms. The velocity data for the 2.12~\mic\
line centres (Table~\ref{t:ratio}) show that motions in the gas must
be small; differences of at most 10~\kms\ are seen between the
different regions, with the exception of P~18-NW (Section~\ref{s:bob}).



\subsection[]{Line Ratios}\label{s:ratios}
\subsubsection{Observed Parameter Space}

For a dense PDR, where self-shielding is occurring, H$_{2}$ exists at
optical depths A$_{V} < 1$ from the front surface of the cloud.  Then,
for a sufficiently strong FUV radiation field, the gas temperature can
reach $\sim 1000$~K or higher, enabling collisions to redistribute the
populations in the H$_{2}$ energy levels. The $(1-0)/(2-1)~S(1)$ line
ratio then increases from the pure fluorescent value.  It is sensitive
to both the density and strength of the FUV field, and so can be used
to constrain these parameters in a source.  This is different from
shock-excited gas where it has been shown empirically that there is
little variation in this line ratio both among and within shocked
sources (\eg\ Burton \etal\ 1989).  In particular, dense clumps of
H$_{2}$ inside the PDRs may have the ${\rm v}=1$ level thermalised,
while being surrounded by lower density gas exhibiting pure
fluorescent line ratios. Such clumps would signal their presence by
elevated levels of the $(1-0)/(2-1)~S(1)$ ratio.  This could then be used
to probe the structure of the clouds.

\begin{figure}
\vskip 9cm
\includegraphics{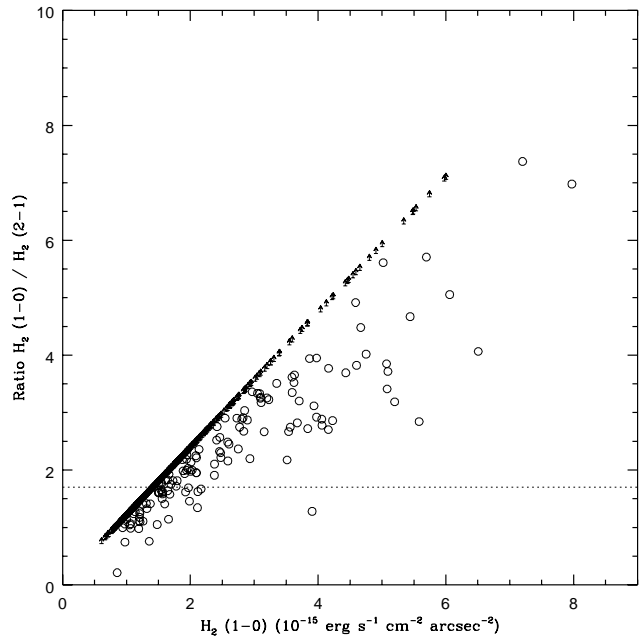}
\caption{Plot of the \H2\ $(1-0)/(2-1)~S(1)$ line ratio against $1-0$
line flux in P~18. Open circles are points which have ${\rm S/N} > 3$ in 
both lines, while the arrows denote points for which the S/N of the $1-0$
line is $> 3$, but the S/N of the $2-1$ line is $< 3$; thus these are lower 
limits in the ratio. The dashed horizontal
line marks a ratio of 1.7, the value expected for purely fluorescent emission.
Both images were smoothed by a Gaussian function with $\sigma=0.3$ before
the S/N criterion was applied.}
\label{f:ratio}
\end{figure}

\begin{figure*}
\vskip 14cm
\includegraphics{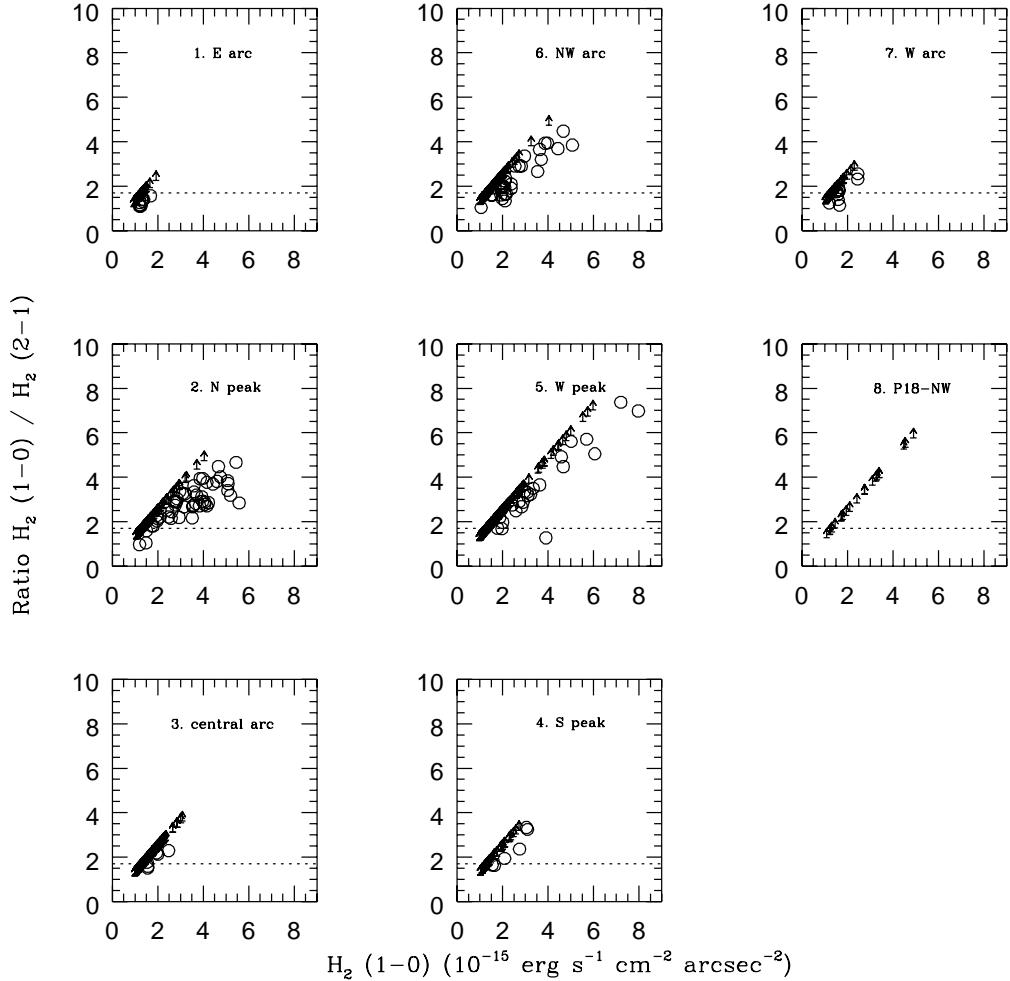}
\caption{Same as Figure~\protect{\ref{f:ratio}}, but for the specific
regions of P~18 outlined in Figure~\protect{\ref{f:h2r}} and listed in
Table~\protect{\ref{t:ratio}}.} 
\label{f:ratio8}
\end{figure*}

We have thus sought to determine the fine-structure in the spatial
variation of the $(1-0)/(2-1)~S(1)$ ratio. We have registered the two
images and applied a light Gaussian smooth ($\sigma = 0.3$). In Figure
\ref{f:ratio} we plot the line ratio so derived against the
$(1-0)~S(1)$ line intensity.
Two sets of symbols are shown in Fig~\ref{f:ratio}; open circles for data
points where both the $1-0$ and $2-1~S(1)$ lines have ${\rm S/N} > 3$, and
lower limits for pixels where the S/N for the $2-1~S(1)$ is $< 3$ (calculated
as though its flux were $3 \sigma$). The line ratio rises approximately
linearly as the $1-0~S(1)$ line flux increases, to a maximum line ratio
of $\sim 8$, but the dispersion in line ratio increases also. 
Similar plots for each of the eight distinct regions marked on
Figure~\ref{f:h2r} are presented in Figure~\ref{f:ratio8} (which
also illustrates the effect of varying the FUV field) and
indicate that most of this spread in line ratio for a given
$1-0~S(1)$ intensity is intrinsic to regions 2 and 6 (\ie\ the northern
regions of P~18). Our interpretation of such trends must be
guarded however, owing to the large number of lower limits (non-detections
in the $2-1~S(1)$ line); high ratios for low $1-0~S(1)$ fluxes
cannot be excluded by our data. Nevertheless, line ratios consistent
with pure fluorescence become progressively rarer as the $1-0~S(1)$
line flux increases. We now seek to explain the apparent variations in
line ratio with emission line flux.

\subsubsection{Interpretation}

In pure fluorescent gas we would obtain a line ratio of $\sim1.7$,
independent of the line intensity.  Localised regions of shocked gas
would also be evident by a ratio of $\sim 10$ for a range of line
intensities (see Brand \etal\ 1989). However we see neither such
behaviour in P~18 (with the exception of P~18-NW, which we argue in
Section~\ref{s:bob} to be physically unassociated with the
other molecular hydrogen emission regions).

Qualitatively the increase in line ratio can be understood as follows.
At low density and FUV fields, the $(1-0)/(2-1)~S(1)$ ratio is purely
fluorescent.
However for a fixed FUV field, H$_{2}$ of sufficiently high density will be
self-shielded, producing a thermal contribution to the $1-0~S(1)$ line.
Its flux will rise as the density does, while the $2-1~S(1)$ line
remains purely fluorescent.  Thus we expect both the strength of the
$1-0~S(1)$ line and the ratio with the $2-1~S(1)$ line to rise as the
density rises.

This line ratio variation can also be understood quantitatively
through the PDR models of BHT\@. BHT derive a number of scaling
relations depending on (i) whether n/G$_{0}$ is less than, or greater
than 40~\cmt\ (determining if H$_{2}$ self-shielding occurs), and
(ii) whether n is less than, or greater than n$_{\rm crit}$, the density where
collisional de-excitation equals the radiative decay rate from a
level. The line intensity can be divided into a UV--pumped
($\rm I_{pump}$) and a thermal ($\rm I_{thermal}$) part, as follows:

When n/G$_{0} \preceq 40$~\cmt\ (self-shielding unimportant):

\begin{equation}
 {\rm I_{pump}} \propto \left\{ \begin{array}{ll}
                      {\rm n}    &  {\rm (n < n_{crit})}  \\
                  {\rm n_{crit}} &  {\rm (n > n_{crit})} \label{eq:pumpssu}
                                  \end{array}
                     \right.
\end{equation}

When  n/G$_{0} \succeq 40$~\cmt\ (self-shielding important):

\begin{equation}
 {\rm I_{pump}} \propto \left\{ \begin{array}{ll}
                    {\rm G_0}    &  {\rm (n < n_{crit})}  \\
          {\rm G_0 n_{crit} / n} &  {\rm (n > n_{crit})} \label{eq:pumpssi}
                                  \end{array}
                          \right.
\end{equation}

The thermal contribution is given by:

\begin{equation}
 {\rm I_{thermal}} \propto \left\{ \begin{array}{ll}
       {\rm n N_2 \gamma}       &  {\rm (n < n_{crit})} \\
     {\rm N_2 e^{-E/kT} / Z(T)} &  {\rm (n > n_{crit})} \label{eq:thermal}
                                     \end{array}
                             \right.  
\end{equation}

\noindent
where N$_{2}$ is the column of warm H$_{2}$ (\ie\ at A$_{V} < 1$) with
temperature $\sim {\rm T}$, E is the energy of the level of interest,
$\gamma$ is the collisional excitation rate from the ground state, and
Z(T) is the H$_{2}$ partition function.

For the bulk of the gas, with density ${\rm n} \sim (2-4) \times
10^4$~\cmt\ and G$_{0} \sim 10^{3-4}$, we have both $\rm n < n_{crit}$
and n/G$_{0} < 40$~\cmt.  The line intensity is then simply
proportional to the product of the density and the filling factor, f.
To within a factor of 3, for $\rm n \prec 10^5$~\cmt, ${\rm
I}_{1-0~S(1)} \sim 3\, {\rm f}\, {\rm n}_4\, \times
10^{-16}$~ergs~\cmtwo~\sone~\arc, where n$_{4}$ is the density in
units of $10^4$~\cmt\ (see equation~\ref{eq:pumpssu}). As discussed in
\S~\ref{s:fluor}, this is consistent with the average line fluxes
observed for filling factors of $1-3$, with $\rm n_4 = 4$, the value
deduced from the geometrical model (see \S~\ref{s:geom}).

The detailed models of BHT show that the thermal contribution to the
line intensities is negligible when G$_{0} = 10^3$, and also when ${\rm
n} \oldleq 10^5$~\cmt\ for G$_{0} = 10^4$.  However for the larger
FUV--field, the thermal contribution to the $1-0~S(1)$ line flux is
$\sim 70\%$ for $\rm n = 10^{6-7}$~\cmt\ (the $2-1~S(1)$ line remains
purely fluorescent).  Thus higher line fluxes and ratios $> 2$ imply
the presence of some gas of at least this density.

For the E and W arcs (see Table~\ref{t:ratio}), where the G$_{0}$
is $\sim 2000$, the line ratio is constant at $\sim 1.7$. In the N and S
peaks and the NW arc, with G$_{0}$ = 5060, 3500, and 3200 respectively,
a thermal
contribution to the line ratio is becoming apparent. At the W peak,
closest to the illuminating source, where G$_{0} = 17800$, both the
$1-0~S(1)$ line intensity and $1-0/2-1~S(1)$ ratio are highest; the
$2-1~S(1)$ line flux density is however, relatively diffuse at this latter
position. On the basis of Figure~\ref{f:ratio8}, it would seem
reasonable that the Central arc, although apparently superimposed
on star `A', must in fact lie at a similar distance from it as the
E and W arcs.

These observations can be understood as follows.  For the arcs, where
the FUV field is low, the emission is purely fluorescent, but the line
ratio is insensitive to density.  At the N and S~peaks, a thermal
contribution would be expected for ${\rm n} \oldgeq 10^6$~\cmt. The line
ratios at the brightest pixels suggest that $\sim 30\%$ of the
$1-0~S(1)$ emission there is thermal, and thus has densities this high.
At the W~peak, self-shielding will occur when $\rm n > 40 G_0 \sim 5
\times 10^5$~\cmt.  The data here are thus sensitive to the presence of
gas denser than this, a density also larger than n$_{\rm crit}$. Ratios
observed are as high as 7, implying that as much as 70\% of the
emitting gas at those positions is at least this dense.

Over the W peak, the specific intensity of the $2-1~S(1)$ line is
relatively weaker, in comparison with that of the $1-0~S(1)$ line,
than at the other peaks excited by star `A'. In this region, the FUV
field G$_{0}$ exceeds $10^{4}$. There is a significant thermal
contribution to the $1-0~S(1)$ flux from dense gas. The $2-1~S(1)$
flux however remains purely fluorescent. Indeed, as seen by
equation~\ref{eq:pumpssi}, its intensity is predicted to decrease as
1/n, consistent with the observation that it gets relatively
weaker. At other peaks, the FUV field is not sufficiently high for us
to expect to observe this behaviour.

The maximum $1-0~S(1)$ line intensity that is purely fluorescent,
arising from gas of average cloud density, occurs when the filling
factor is $\sim 3$, and the flux density is $\sim 4 \times
10^{-16}$~ergs~\cmtwo~\sone~\arc\ for the spherical shell geometry
we have modeled.  For higher line fluxes a thermal
component is present, and the line ratio rises above the pure
fluorescent value.  The thermal component arises from gas of density
at least $10^6$~\cmt, and both the highest line fluxes and line ratios
occur when the filling factor of this dense gas is greatest.

The spread of observed line ratios between the minimum and maximum
values for a fixed $1-0~S(1)$ line flux can be understood by a
variation of the relative filling factors for the low density and high
density gas.  The lowest ratio occurs when the proportion of low
density gas is highest and rises, for a fixed flux level, as both the
fraction of low density gas falls and that of the high density gas
rises.  Thus data with sufficiently high S/N to accurately map the
line ratio variation from pixel to pixel could be used to determine
the micro-structure of the clumping. The line ratio map
(Figure~\ref{f:h2r}), for positions where G$_0 \succeq 3000$, is
effectively showing the relative amounts of high and low density gas.

We thus conclude that the variations in the line ratio are consistent
with the PDR models, self-shielding of gas occurring in its densest
parts, with thermal contributions evident when exposed to the highest
radiation fields. Localised regions where the density is more than ten
times the average exist throughout the molecular cloud. The highest
line fluxes and line ratios occur together, where both the filling
factor of low and high density gas is greatest, and changes in line
ratio from position to position reflect changes in the relative filling
factors of these components.

\subsection[]{P~18-NW}\label{s:bob}

The isolated flux peak to the northwest (designated by us as P~18-NW)
however, shows no evidence for fluorescent line emission. The
$2-1~S(1)$ line here is barely detected, and the $1-0/2-1~S(1)$ line
ratio is clearly thermal in nature, similar to shock-excited
values. P~18-NW is too far from star `A' for collisional
fluorescence to be significant, as G$_0 < 500$.  We suggest that there
is a localised source of heating, possibly shocking the gas.

The simultaneous velocity coverage of UNSWIRF offers at least one
clue. We notice that the line centre velocity in P~18-NW is also quite
different to the rest of P~18. On account of its redshift of
$+14$~\kms relative to the central arc, we speculate that P~18-NW
might in fact be associated in some way with the CO outflow source
observed by Fukui (1989) and by Felli, Palagi \& Tofani (1992).
Although Fukui identified the outflow with the position of star `A',
the 3~arcmin beam used for the CO observations is much larger than the
45~arcsec separation between star `A' and P~18-NW. The {\em IRAS}
point source discussed by L\'{o}pez et~al. (1988) and by Felli
et~al. (1992), namely IRAS $06572-0742$, has a bolometric luminosity
$\rm \sim 1500\,L_{\odot}$ and colours suggestive of an ultra-compact
H\,{\sc ii}~region, though neither H$_2$O (Felli \etal\ 1992) nor
methanol (Walsh \etal\ 1997) masers have been detected from it. While
the {\em IRAS} positional uncertainties do not preclude an association
of P~18-NW with the IRAS source, it is perhaps more likely that the
H$_{2}$ line emission observed from P~18-NW is shock-excited by the
interaction of an outflow, from an embedded source in P~18, with the
ambient cloud material.  More detailed CO observations are needed to
clarify the influence of any outflow activity on the excitation and
environment of P~18.

\section{Conclusions}\label{s:conc}

The reflection nebula Parsamyan~18 has been imaged in both the 2.122
and 2.248~\mic\ emission lines of molecular hydrogen using the UNSW
Infrared Fabry-Perot on the AAT, allowing us to study variations in
their ratio within the nebula.  The emission-line morphology appears
as a pair of overlapping arcs, consistent with incomplete PDR shells
irradiated by a single early B-type star. We believe we can identify
this star in our continuum maps, which contradicts previous claims
that the stellar-like sources in P~18 are `windows' in a circumstellar
dust shell, illuminated by a compact \HII~region. Modeling suggests
that the thickness of the fluorescing shells is roughly 25\% of
their radius, with average gas densities of $(2-4)\times10^{4}$~\cmt,
and dust extinction limiting the penetration of the FUV photons. In
addition, some regions of P~18 display line ratios indicative of
collisional fluorescence in the densest regions exposed to the highest
radiation fields.  In general, {\em those regions with the highest
intrinsic $1-0~S(1)$ intensities also have the highest
$(1-0)/(2-1)~S(1)$ line ratios}, a result of this process. Within
these regions, large ranges in the observed line ratio can be
understood as arising from the underlying variations in the relative
filling factors of high- and low-density gas. We have discovered an
isolated knot of shock-excited \H2\ gas just outside the nebula, which
we argue on kinematical grounds is more likely to be associated with
the CO outflow source reported in this region. This study highlights
the advantages of an imaging Fabry-Perot system over traditional
long-slit spectroscopy for emission-line imaging, and for
understanding the excitation mechanism of molecular hydrogen in
particular.

\section*{Acknowledgments}

We acknowledge useful discussions with Peter Brand and Antonio
Chrysostomou, and suggestions from the referee for improving the
paper. We thank Yin-Sheng Sun for valuable assistance in calibrating
UNSWIRF. S.D.R. acknowledges the receipt of a UNSW Vice-Chancellor's
Postdoctoral Fellowship. UNSWIRF was funded by a grant from the
Australian Research Council.

\bsp

\label{lastpage}


\begin{thebibliography}{99}
\bibitem{daa} Allen D. A. et~al., 1993, PASA, 10, 298
\bibitem{bd}  Black J. H., Dalgarno A., 1976, ApJ, 203, 132
\bibitem{bvd} Black J. H., van Dishoeck E. F., 1987, ApJ, 322, 412
\bibitem{btg} Brand P. W. J. L., Toner M. P., Geballe T. R., Webster A. S.,
              Williams P.M., Burton M. G., 1989, MNRAS, 236, 929
\bibitem{bgw} Burton M. G., Brand P. W. J. L., Geballe T. R., Webster A. S.,
              1989, MNRAS, 236, 409
\bibitem{mgb} Burton M. G., Geballe T. R., Brand P. W. J. L.,
              Moorhouse A., 1990, ApJ, 352, 625
\bibitem{bht} Burton M. G., Hollenbach D. J., Tielens A. G. G. M., 1990,
              ApJ, 365, 620 (BHT)
\bibitem{c84} Cohen M., 1974, PASP, 86, 813
\bibitem{fpt} Felli M., Palagi F. Tofani G., 1992, A\&A, 255, 293
\bibitem{f89} Fukui Y., 1989, in Low Mass Star Formation and Pre-Main
              Sequence Objects, ed. B. Reipurth (Garching: ESO), 95
\bibitem{g87} Gatley I., Hasegawa T., Suzuki H., Garden R., Brand P.,
              Lightfoot J., Glencross W., Okuda H., Nagata T., 1987,
              ApJ, 318, L73
\bibitem{h68} Habing H. J., 1968 Bull. Astr. Inst. Netherlands, 19, 421
\bibitem{hay} Hayashi M., Hasegawa T., Gatley I., Garden R., Kaifu, N.,
              1985, MNRAS, 215, 31P
\bibitem{hl}  Hilton J., Lahulla J. F., 1995, A\&AS, 113, 325
\bibitem{h94} Hodapp K.-W., 1994, ApJS, 94, 615
\bibitem{lop} L\'{o}pez J. A., Roth M., Friedman S. D., Rodr\'{i}guez L. F.,
              1988, Rev. Mexicana Astron. Astrofis., 16, 99
\bibitem{r97} Ryder S. D., Sun Y.-S., Storey J. W. V., Ashley M. C. B.,
              Burton M. G., Allen L. E., 1997, in preparation
\bibitem{s86} Sellgren K., 1986, ApJ, 305, 399
\bibitem{swa} Sellgren K., Werner M. W., Allamandola L. J., 1996, ApJS, 102,
              369 
\bibitem{std} Sternberg A., Dalgarno A., 1989, ApJ, 338, 197 
\bibitem{stu} Stutzki J., Stacey G. J., Genzel R., Graf U. U., Harris A. I.,
              Jaffe D. T., Lugten J. B., Poglitsch A., 1990, in
              Submillimetre Astronomy, eds. G. D. Watt \& A. S. Webster
              (Dordrecht: Kluwer), 269
\bibitem{tho} Tielens A. G. G. M. Hollenbach D. J., 1985, ApJ, 291, 722
\bibitem{ajw} Walsh A. J., Hyland A. R., Robinson G., Burton M.G., 1997,
              MNRAS, in press
\end{thebibliography}
\end{document}